\documentclass[twocolumn,english,aps,prb,10pt,superscriptaddress,showpacs]{revtex4}
\usepackage[T1]{fontenc}
\usepackage{amsmath,graphicx,amssymb,epsfig,babel,dsfont,mathrsfs,color}

\begin{document}

\title{Radiative cooling induced by time-symmetry breaking in periodically-driven systems}

\author{Riccardo Messina}
\affiliation{Laboratoire Charles Fabry, UMR 8501, Institut d'Optique, CNRS, Universit\'{e} Paris-Saclay, 2 Avenue Augustin Fresnel, 91127 Palaiseau Cedex, France.}

\author{Annika Ott}
\affiliation{Institut f\"{u}r Physik, Carl von Ossietzky Universit\"{a}t, D-26111 Oldenburg, Germany.}

\author{Christoph Kathmann}
\affiliation{Institut f\"{u}r Physik, Carl von Ossietzky Universit\"{a}t, D-26111 Oldenburg, Germany.}

\author{Svend-Age Biehs}
\affiliation{Institut f\"{u}r Physik, Carl von Ossietzky Universit\"{a}t, D-26111 Oldenburg, Germany.}

\author{Philippe Ben-Abdallah}
\email{pba@institutoptique.fr} 
\affiliation{Laboratoire Charles Fabry, UMR 8501, Institut d'Optique, CNRS, Universit\'{e} Paris-Saclay, 2 Avenue Augustin Fresnel, 91127 Palaiseau Cedex, France.}

\date{\today}


\begin{abstract}
	We theoretically study the thermal relaxation of many-body systems under the action of oscillating external fields. When the magnitude or the orientation of a field is modulated around values where the pairwise heat-exchange conductances depend non-linearly on this field, we demonstrate that the time symmetry is broken during the evolution of temperatures over a modulation cycle. We predict that this asymmetry enables a pumping of heat which can be used to cool down faster the system. This effect is illustrated through different magneto-optical systems under the action of an oscillating magnetic field. 
\end{abstract}

\maketitle

Understanding and controlling the thermal relaxation of systems out of thermal equilibrium is a major problem in physics both on the fundamental and applied point of view. The dynamic control of radiative heat exchanges between solids or between a solid and its surrounding environment plays an important role in this process. By modulating some intensive quantities such as the temperature or the chemical potential in one of these solids a supplementary flux is generated, giving an additional contribution to the primary flux induced by the mean gradient. This shuttling effect~\cite{Latella2018,Messina} can be used to enhance the radiative heat transfer between two bodies or to thermally insulate them. However this mechanism is intrinsically limited by the system's thermal inertia so that the system can only be adiabatically cooled down. In 2019 another mechanism has been proposed to control the magnitude of radiative heat exchanged in thermal photonic systems which highlight topological singularities~\cite{Li}. In such systems an adiabatic radiative pumping effect has been demonstrated. Nevertheless, this pumping can only be observed in systems which present such singularities. In a recent work~\cite{Fan1} Buddhiraju et al. studied the possibility to cool down solids whose refractive index undergoes temporal modulation. When the dielectric permittivity of the system undergoes a sufficiently fast temporal modulation, light emission and absorption become radically different phenomena. In the present work we investigate the relaxation dynamics of systems driven by a time-dependent external field $\mathbf{F}(t)$ without breaking the emission\,-\,absorption symmetry. We predict the possible existence of a pumping effect which can be used to control the relaxation dynamics and in specific conditions to accelerate the relaxation process, paving thus the way for an active thermal management at the nanoscale and for innovative cooling applications.

To address this problem, let us start by analyzing the relaxation dynamics of a single system interacting with its environment through a heat-exchange conductance $G[\mathbf{F}(t)]$, depending on an external field $\mathbf{F}$ which can be controlled with respect to time. If we assume that this system is initially prepared at a temperature higher than its environment, it will naturally cool down toward its equilibrium (i.e. the environmental temperature) according to Newton's law of cooling
\begin{equation}
C\frac{dT}{dt}=G[\mathbf{F}(t)] (T_b-T),\label{Newton}
\end{equation}
where $C=\rho C_hV$ denotes the product of the system density, specific heat capacity and volume, while $T_b$ stands for the environmental temperature. The solution of Eq.~\eqref{Newton} reads
\begin{equation}
T(t)=T_b+(T(0)-T_b)\exp\Bigl[-\frac{1}{C}\int_{0}^{t}G[\mathbf{F}(\tau)] d\tau\Bigr].\label{Newton_solution}
\end{equation}
If we consider an external field $\mathbf{F}(t)=\mathbf{F}_0+\Delta\mathbf{F}\sin(\omega t)$, oscillating at frequency $\omega$ around a constant value $\mathbf{F}_0$, the conductance can be expanded around $\mathbf{F}_0$ as follows
\begin{equation}
\begin{split}
G[\mathbf{F}(t)] &= G(\mathbf{F}_0)+\nabla_F G(\mathbf{F}_0)\cdot\Delta\mathbf{F}\sin(\omega t)\\
&\,+\frac{1}{2}{\Delta\mathbf{F}}^t\cdot\mathbb{H}_G(\mathbf{F}_0)
\cdot\Delta\mathbf{F} \sin^2(\omega t)+\dots\label{G_expansion}
\end{split}
\end{equation}
where $\mathbb{H}_G(\mathbf{F}_0)$ is the Hessian matrix associated with the conductance around the static field $\mathbf{F}_0$. Using this expansion up to the second order the exponential damping term in Eq.~\eqref{Newton_solution} reads, over an oscillation period $T=2\pi/\omega$,
\begin{equation}
\exp\biggl[-\frac{1}{C}\int_{0}^{T} G[\mathbf{F}(\tau)] d\tau\biggr] = \exp\biggl[-\frac{T}{C} \biggl( G[\mathbf{F}_0] + \mathscr{G} \biggr) \biggr]
\label{damping}\end{equation}
introducing the dynamical damping or pumping factor $\mathscr{G} = \frac{1}{4}\Delta\mathbf{F}^t\cdot\mathbb{H}_G(\mathbf{F}_0)\cdot\Delta\mathbf{F}$.
As the first term of RHS corresponds to the usual damping term for a relaxation process driven by a constant heat-exchange conductance, the second term describes the effect on this damping due to the superimposed field oscillation. We see that, depending on the sign of the curvature of $G$ with respect to $\mathbf{F}$, the external field can either accelerate (i.e. $\mathscr{G} > 0$) or decelerate (i.e. $\mathscr{G} < 0$) the cooling process.

This simple model allows a straightforward generalization to $N$-body systems. To this aim, we consider a system made of $N$ objects having $C_i=\rho_iC_{h,i}V_i$ and temperatures $T_i$, subject to pairwise interactions and coupled to a common external bath at temperature $T_b$. The thermal state $(T_1,...,T_N)$ of this system thus evolves according to
\begin{equation}
C_i\frac{dT_i}{dt}=\wp_{i}+\wp_{bi},
\end{equation}
where $\wp_{i}=\sum_{j\neq i}\wp_{ji}$ is the power exchanged between body $i$ and all others bodies, $\wp_{ji}=\wp_{ji}(T_1,...,T_N;T_b)$ being the power exchanged between body $j$ and body $i$, while $\wp_{pi}=\wp_{bi}(T_1,...,T_N;T_b)$ denotes the power the environment exchanges with body $i$. Close to thermal equilibrium $\mathbf{T}_\mathrm{eq}=(T_b,\dots,T_b)$ the system dynamics satisfies, in terms of thermal conductances $G_{ji}=\frac{\partial\wp_{ji}}{\partial T}$,
\begin{equation}
C_i\frac{dT_i}{dt}=\underset{j\neq i}{\sum}G_{ji}(T_j-T_i)+G_{bi}(T_b-T_i),
\end{equation}
which is the straightforward generalization of Newton's heat law. This equation can be recast in a matrix form as
\begin{equation}
 \mathds{C}\frac{d\mathbf{T}}{dt} = \mathds{G}[\mathbf{F}(t)]\mathbf{T}(t)+\mathbf{S}.
\label{Eq:diff2}
\end{equation}
where $\mathbf{T}=(T_{1},\dots,T_{N})$ is the system thermal state, $\mathds{C}={\rm diag}(C_1,\dots,C_N)$ is the diagonal inertia matrix, $\mathbf{S}=T_b\,\mathrm{diag}(G_{b1},\dots,G_{bN})$ is the source term, while $\mathds{G}$ is the general conductance matrix with components
\begin{equation}
 \mathds{G}_{ij}= -\biggl(\sum_{k\neq i} G_{ki}+G_{bi}\biggr)\delta_{ij}+ (1-\delta_{ij})G_{ji}.
\label{mat_cond1}
\end{equation}
As for a single object, if we assume that the thermal conductances (including interaction conductances with the thermal bath) depend on an external (not necessarily uniform) field we can write the Taylor expansion 
\begin{equation}
\begin{split}
G_{ji}[\mathbf{F}(t)] &= G_{ji}(\mathbf{F}_0)+\nabla_F G_{ji}(\mathbf{F}_0)\Delta\mathbf{F}\sin(\omega t)\\
&\,+\frac{1}{2}{\Delta\mathbf{F}}^t\cdot\mathbb{H}_{G_{ji}}(\mathbf{F}_0)
\cdot\Delta\mathbf{F}\sin^2(\omega t)+\dots\label{G_ij_expansion}
\end{split}
\end{equation}
for these conductances with respect to this field. The time average of the net power $\wp_{i,\mathrm{net}}=\underset{j\neq i}{\sum}G_{ji}(T_j-T_i)$ received by each body during one oscillating period of $\mathbf{F}$ is
\begin{equation}
\langle\wp_{i,\mathrm{net}}\rangle= \underset{j\neq i}{\sum}\langle G_{ji}\rangle(T_j-T_i)=\wp_{i,\mathrm{net}}\mid_{\mathbf{F}_0}+\Delta\wp_{i,\mathrm{net}},
\label{average_power}
\end{equation}
where $\wp_{i,\mathrm{net}}\mid_{\mathbf{F}_0}=\underset{j\neq i}{\sum} G_{ji}(\mathbf{F}_0)(T_j-T_i)$ is the net power received by the body $i$ under a static field $\mathbf{F}_0$ and
\begin{equation}
\Delta\wp_{i,\mathrm{net}}=\underset{j\neq i}{\sum} \mathscr{G}_{ji} (T_j-T_i)
\label{delta_power}
\end{equation}
is the extra-power which superimposes to this power during one oscillating period of the field $\mathbf{F}$, where the pumping factor for each interacting channel is this time defined by $\mathscr{G}_{ji} = \frac{1}{4}{\Delta\mathbf{F}}^t\cdot\mathbb{H}_{G_{ji}}(\mathbf{F}_0)\cdot\Delta\mathbf{F}$. Here we assumed that the field oscillation frequency is such that variations of the body temperatures $T_i$ are negligible during a single oscillation period. Under these conditions, we see that the cooling effect depends on the convexity of the pairwise conductances $G_{ji}$ .

Unlike the case of a simple object, it is not so direct to assess the impact of this extra-power on the relaxation dynamics. Nevertheless, this effect can be rigorously investigated using the general framework of Floquet theory~\cite{Floquet,Hale}. Introducing the fundamental matrix $\mathds{X}(t)$ of the differential system \eqref{Eq:diff2}, the time evolution of the thermal state reads
\begin{equation}
\mathbf{T}(t)=\mathds{X}(t)\mathbf{T}(0)+\mathds{X}(t)\int_{0}^{t}\mathds{X}^{-1}(\tau) \mathds{C}^{-1}\mathbf{S}(\tau)d\tau.\label{Nbody_solution}
\end{equation}
For a $T$-periodic modulation of $\mathbf{F}$, the fundamental matrix takes the form $\mathds{X}(t)=\mathds{Q}(t)e^{t\mathds{R}}$ where $\mathds{Q}(t)$ is a periodic matrix of period $T$ and $\mathds{R}$ is a constant (in general non-diagonal) matrix given by $\mathds{R}=\frac{1}{T}\log[\mathds{X}(T)]$. After one period of modulation we see that $\mathds{X}(t+T)=\mathds{X}(t)e^{T\mathds{R}}$ and the solution becomes
\begin{equation}
\begin{split}
\mathbf{T}(t+T)&=\mathds{X}(t)e^{T\mathds{R}}\mathbf{T}(0)\\
&\,+\mathds{X}(t)e^{T\mathds{R}}\int_{0}^{t+T}\mathds{X}^{-1}(\tau) \mathds{C}^{-1}\mathbf{S}(\tau)d\tau.\label{Nbody_solution2}
\end{split}
\end{equation}
Hence we see that the damping induced by the modulation of $\mathbf{F}$ is entirely described by the matrix $\mathds{R}$. Practically speaking, the fundamental matrix can be easily calculated using the resolvant $\mathcal{R}$ of system \eqref{Eq:diff2} (see Appendix~\ref{AppA} for details). While this approach is needed in the general $N$-body scenario, in the 2-body case the $\mathds{R}$-matrix becomes diagonal and thus the dynamics can be already understood from the curvature of the conductance.

\begin{figure}
	\centering
	\includegraphics[width=0.4\textwidth]{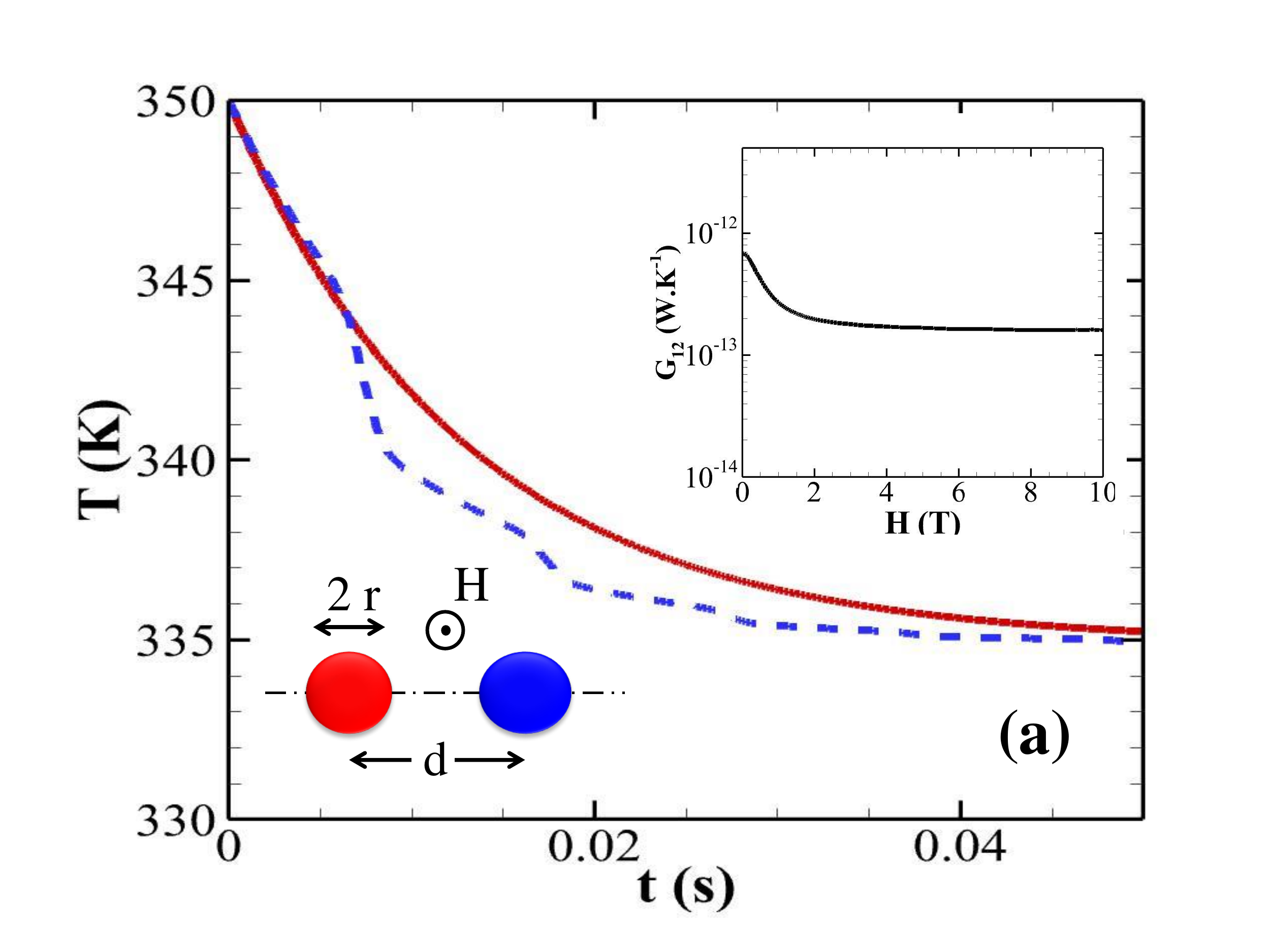}
	\includegraphics[width=0.4\textwidth]{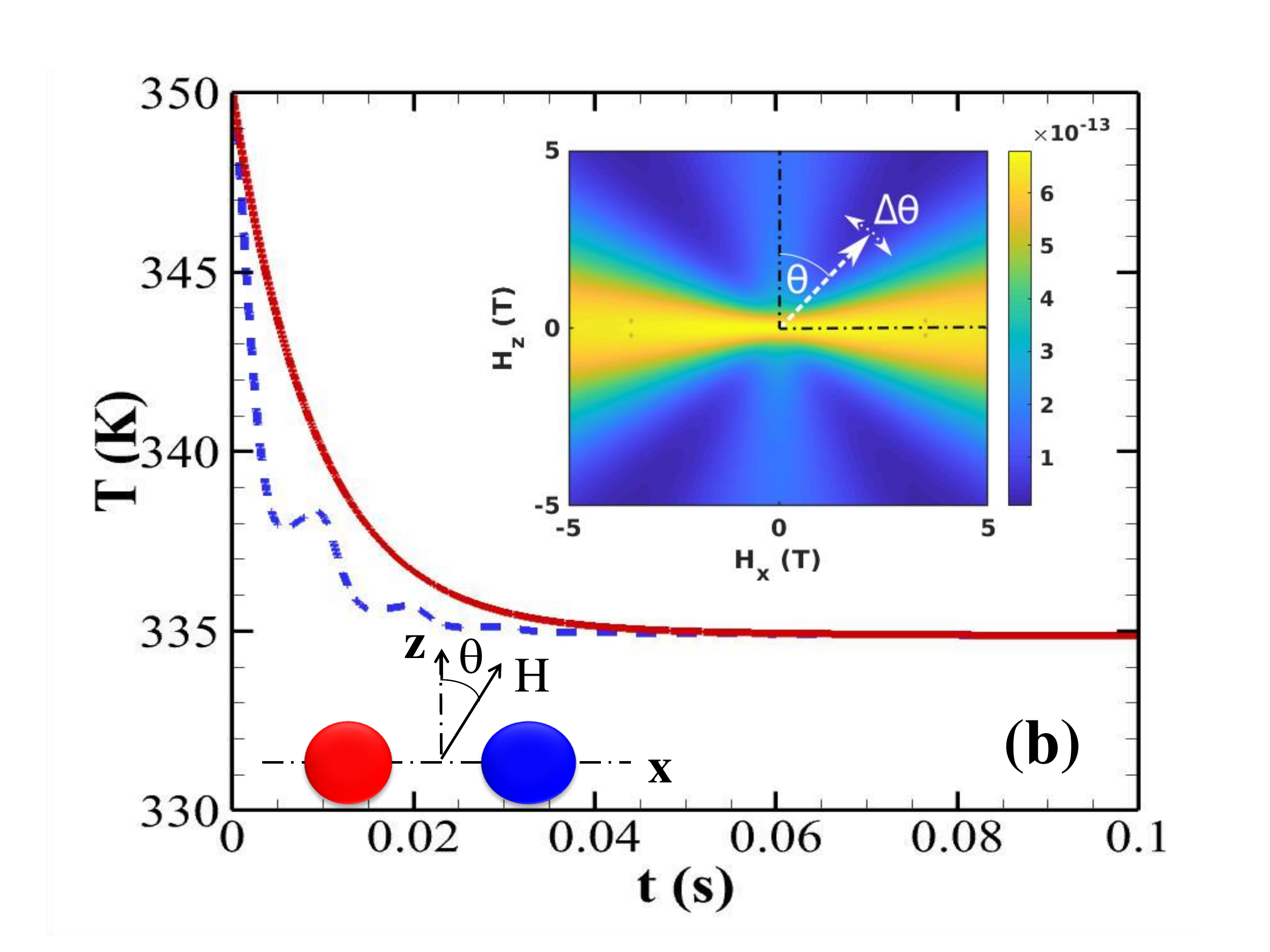}
	\caption{Magnetic cooling of the hot particle in a dimer of InSb nanoparticles ($r=100\,$nm) separated by a distance $d=3R$ and embedded in a thermal bath under the presence of a static magnetic field $H=H_0$ (red solid line) or a sinusoidal field (blue dashed line). (a) The magnetic field is applied in the direction orthogonal to the dimer axis and its magnitude oscillates as $H=H_0+\Delta H \sin(2\pi t/\tau)$ with $H_0=\Delta H=2\,$T. (b) The magnetic field of magnitude $H=2\,$T oscillates sinusoidally in the $(x,z)$ plane around an angle $\theta_0=40\,$degrees over an angular opening $\Delta\theta=10\,$degrees. The initial temperatures and the bath temperature are $T_1(0)=350\,$K, $T_2(0)=320\,$K and $T_b=300\,$K. In both scenarios the oscillation period is $\tau=0.01$\,s. Insets show in (a) the thermal conductances between the two particles as a function of the magnetic field at temperature $T_b$ and in (b) the thermal conductance in the plane $(H_x,H_z)$ at temperature $T_b$.} 
	\label{Fig_1}
\end{figure}

To investigate a concrete situation, we consider systems made of magneto-optical bodies (nanoparticles and films) immersed in a thermal bath and subject to an external magnetic field $\mathbf{H}(t)$ in the $(x,z)$ plan (see Fig.(\ref{Fig_1})). We first consider a couple of $n$-doped InSb nanoparticles of radius $r$, separated by a distance $d$, immersed in a thermal bath at temperature $T_b$ and a magnetic field orthogonal to the axis connecting the particles. In these conditions the permittivity tensor of each particle takes the form~\cite{Moncada}
\begin{equation}
\bar{\bar{\varepsilon}}=\left(\begin{array}{ccc}
\varepsilon_{1}\cos^2\theta+\varepsilon_{3}\sin^2\theta & -i\varepsilon_{2}\cos\theta & \frac{1}{2}\sin(2\theta)(\varepsilon_{1}-\varepsilon_{3})\\
i\varepsilon_{2}\cos\theta & \varepsilon_{1} & i\varepsilon_{2}\sin\theta\\
\frac{1}{2}\sin(2\theta)(\varepsilon_{1}-\varepsilon_{3}) & -i\varepsilon_{2}\sin\theta & \varepsilon_{1}\sin^2\theta+\varepsilon_{3}\cos^2\theta
\end{array}\right),\label{Eq:permittivity}
\end{equation}
with
\begin{equation}
\begin{split}
\varepsilon_{1}(H) &= \varepsilon_\infty\bigg(1+\frac{\omega_L^2-\omega_T^2}{\omega_T^2-\omega^2-i\Gamma\omega}+\frac{\omega_p^2(\omega+i\gamma)}{\omega[\omega_c^2-(\omega+i\gamma)^2]}\bigg),\\
\varepsilon_{2}(H) &= \frac{\varepsilon_\infty\omega_p^2\omega_c}{\omega[(\omega+i\gamma)^2-\omega_c^2]},\\
\varepsilon_{3} &= \varepsilon_\infty\bigg(1+\frac{\omega_L^2-\omega_T^2}{\omega_T^2-\omega^2-i\Gamma\omega}-\frac{\omega_p^2}{\omega(\omega+i\gamma)}\bigg).
\end{split}
\end{equation}
Here $\varepsilon_\infty=15.7$ is the infinite-frequency dielectric constant, $\omega_L=3.62\times10^{13}\,\mathrm{rad\,s}^{-1}$ the longitudinal optical phonon frequency, $\omega_T=3.39\times10^{13}\,\mathrm{rad\,s}^{-1}$ the transverse optical phonon frequency, $\omega_p=(\frac{ne^2}{m^*\varepsilon_0\varepsilon_\infty})^{1/2}$ the plasma frequency of free carriers of density $n=1.68\times10^{17}\,\mathrm{cm}^{-3}$, charge $e$, and effective mass $m^*=1.99\times 10^{-32}\,\mathrm{kg}$, $\varepsilon_0$ the vacuum permittivity, $\Gamma=5.65\times10^{11}\,\mathrm{rad\,s}^{-1}$ the phonon damping constant, $\gamma=3.39\times10^{12}\,\mathrm{rad\,s}^{-1}$ the free carrier damping constant, and $\omega_c=eH/m^*$ the cyclotron frequency. 

In the context of the study of radiative heat transfer in a system of $N$ dipolar particles~\cite{pba5,Messina2013}, it has been shown that the thermalization is a multiscale problem, with a fast thermalization between the particles, followed by a slower collective thermalization toward the environmental temperature. In the case of two InSb nanoparticle we will focus on the interparticle thermalization dynamics and thus consider only the power exchanged in near-field between them, which for two identical nanoparticles this power reads~\cite{Ott,pba_Hall,Fan3,Latella}
\begin{equation}
\wp_{ij} = 3 \int_{0}^{\infty}\frac{d\omega}{2\pi}\, \Bigl[\Theta_i\mathcal{T}_{ij}-\Theta_j \mathcal{T}_{ji}\Bigr],\label{Heatpower}
\end{equation} 
where $\Theta = {\hbar\omega}/[{e^{\frac{\hbar\omega}{k_B T}}-1}]$ is the mean energy of a harmonic oscillator in thermal equilibrium at temperature $T$, whereas $\mathcal{T}_{ij}(\omega)$ denotes the interparticle transmission coefficients (note that in the particular case of a two-body configuration we have $\mathcal{T}_{12}=\mathcal{T}_{21}$ \cite{Latella}). When the particles are small enough compared with their thermal wavelength they can be modeled as simple radiating electrical dipoles. In this case the transmission coefficient is given by~\cite{Cuevas,Cuevas2}
\begin{equation}
	\mathcal{T}_{ij}(\omega)=\frac{4}{3}\Bigl(\frac{\omega}{c}\Bigr)^4\text{Im}\text{Tr}\Bigl[\bar{\bar{\alpha}}_j\mathds{G}_{ji}\frac{1}{2}(\bar{\bar{\alpha}}_i-\bar{\bar{\alpha}}_i^{\dagger})\mathds{G}_{ji}^{\dagger}\Bigr],
\end{equation}
where $\bar{\bar{\alpha}}_i$ denotes the polarizability tensor of particle $i$ and $\mathds{G}_{ij}$ is the dyadic Green tensor between particles $i$ and $j$ in the system of coupled dipoles~\cite{Purcell}. In the infrared and far-infrared the radiation corrections~\cite{Moncada,Albaladejo} can be negected and the polarizability can be simply expressed for subwavelength nanoparticles embedded in vacuum as
\begin{equation}
\bar{\bar{\alpha}}_i(\omega) = 4\pi r^3\big(\bar{\bar{\varepsilon}}_i- \mathds{1}\big)\big(\bar{\bar{\varepsilon}}_i+2\mathds{1}\big)^{-1},\label{Eq:Polarizability2}
\end{equation} 
$\bar{\bar{\varepsilon}}_i$ being the permittivity tensor associated with particle $i$. Finally, Eq. \eqref{Heatpower} allows us to define the thermal conductance $G_{12}=G_{21}=\lim_{|\Delta T|\to0} \wp_{12}/|\Delta T|$, where $\Delta T=T_1-T_2$. In a two-body system this conductance simply reads
\begin{equation}
G_{12} = 3 \int_{0}^{\infty}\frac{d\omega}{2\pi}\frac{\partial\Theta}{\partial T}\mathcal{T}_{12}(\omega) \label{cond_dipole}
\end{equation}
and its curvature with respect to the external field defines the pumping factor $\mathscr{G}$.

In Fig.~\ref{Fig_1} we show the cooling dynamics of a hot nanoparticle in a dimer of InSb particles initialy prepared at two different temperatures in the presence of both a static and a modulated magnetic fields at ambient temperature, when either the magnitude [Fig.~\ref{Fig_1}(a)] or its orientation [Fig.~\ref{Fig_1}(b)] is modulated. In the insets of Fig.~\ref{Fig_1} are plotted the heat-exchange conductances between the two particles with respect to the magnetic field intensity and with respect to its orientation. It is straightforward to see from these curves that the thermal conductance does not follow a symmetric evolution when the magnetic field is periodically modulated. When this modulation takes place in a region where the conductance is convex, we see that the field oscillation gives rise, according to the previous theoretical aguments, to a pumping effect which accelerates the cooling of the particle compared to the situation where a static field is applied. It is worthwhile to note that in the dimer this pumping effect only exists up to the thermalization of two particles at the same temperature. Indeed, for longer times the two particles relax toward the temperature of the surrounding bath with a far-field heat exchange conductance $G_{bi}$ which is typically several orders of magnitude smaller than the conductance $G_{12}$. Notice that an exact calculation of transmission coefficients from the exact solver SCUFF-EM~\cite{SCUFF1,SCUFF2,Chen} of electromagnetic scattering problem has demonstrated that the role played by the multipoles on the heat transfer is negligible for the particle sizes and distances considered in the present study.

\begin{figure}
	\centering
	\includegraphics[width=0.47\textwidth]{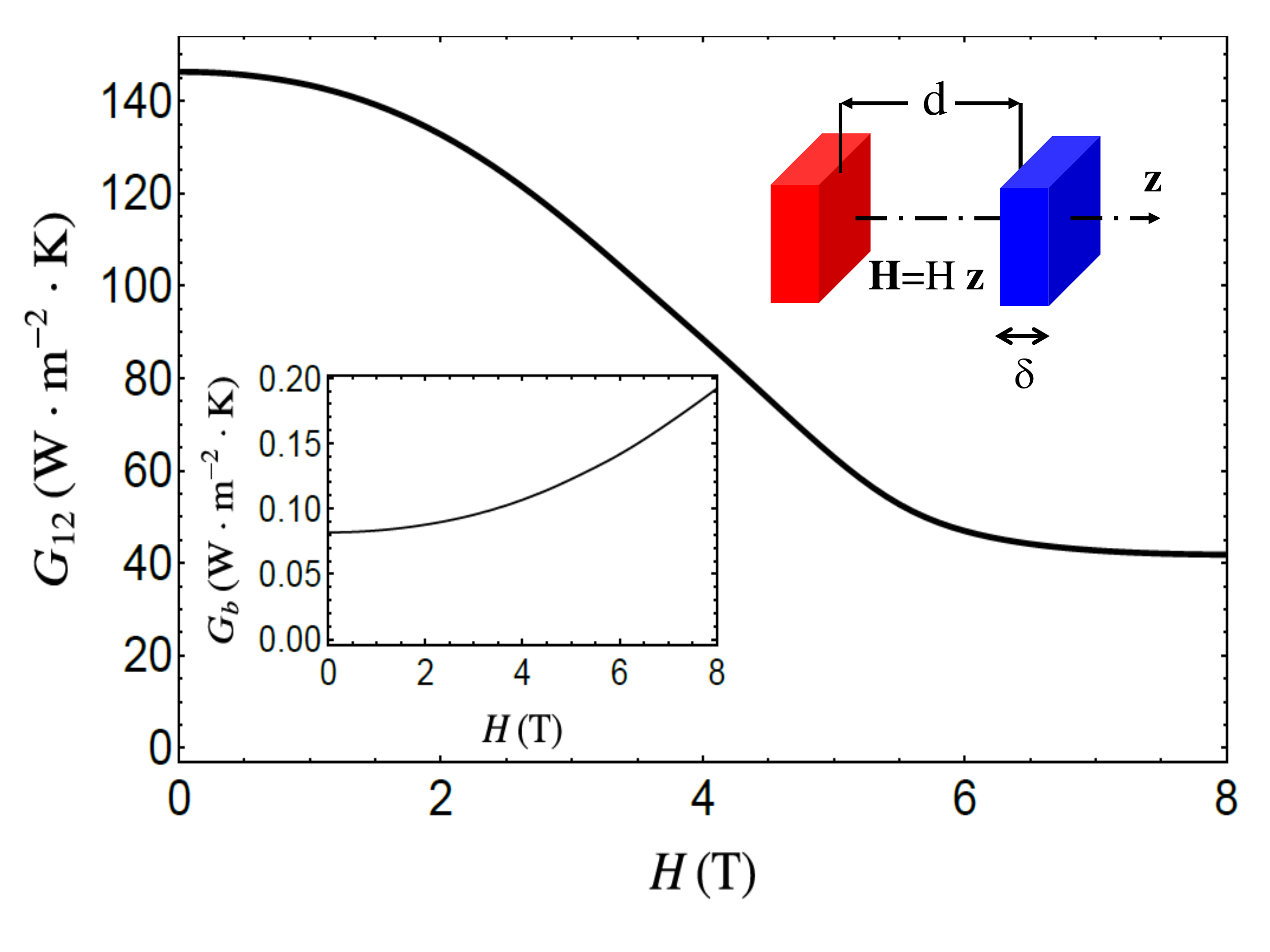}
	\caption{Conductances $G_{12}$ between the two InSb films and $G_b$ between each film and the environment, as a function of the applied magnetic field $H$ along the z-axis and at temperature $T=300\:K$, for films of thickness $\delta=1\,\mu m$ separated by a gap of thickness $d=100\,$nm.}
	\label{Fig_filmconductance}
\end{figure}

We now focus on the case of two planar slabs separated by a gap of thickness $d$ and orthogonal to the $z$ axis, along which the magnetic field $H$ is applied (Fig.~\ref{Fig_filmconductance}). In this case, the thermal conductance between the two films reads
\begin{equation}
G_{12}=\int_{0}^{\infty}\frac{d\omega}{2\pi}\frac{\partial\Theta}{\partial T}\int\frac{d^2\mathbf{k}}{(2\pi)^2}\mathcal{T}_{12}(\omega,k),\label{cond_film}
\end{equation}
with the transmission coefficient ($\omega$ and $k$ are implicit)~\cite{Fan2020}
\begin{equation}
\mathcal{T}_{12}=\begin{cases}
\mathrm{Tr}\bigl[\mathbb{U}^{(21)}(\mathds{1}-\mathbb{R}_2\mathbb{R}_2^\dag-\mathbb{T}_2\mathbb{T}_2^\dag)\\
\quad\times \mathbb{U}^{(21)\dag}(\mathds{1}-\mathbb{R}_1^\dag\mathbb{R}_1-\mathbb{T}_1^\dag\mathbb{T}_1)\bigr], & \omega>ck,\\
\mathrm{Tr}\bigl[\mathbb{U}^{(21)}(\mathbb{R}_2-\mathbb{R}_2^\dag)\mathbb{U}^{(21)\dag}(\mathbb{R}_1^\dag-\mathbb{R}_1)\bigr], & \omega<ck,
\end{cases}
\label{trans_film}\end{equation}
where the reflection and transmission matrices $\mathbb{R}_i$ and $\mathbb{T}_i$ have the following structure
\begin{equation}
\mathds{R}_i=\begin{pmatrix}
r_{i,ss} & r_{i,sp}\\
r_{i,ps} & r_{i,pp}
\end{pmatrix}, \qquad \mathds{T}_i=\begin{pmatrix}
t_{i,ss} & t_{i,sp}\\
t_{i,ps} & t_{i,pp}
\end{pmatrix}
\end{equation}
written in term of Fresnel coefficients $r_{i,uw}$ and $t_{i,uw}$ for both polarization states $u,w=s,p$. The explicit expression of $\mathbb{R}_i$ and $\mathbb{T}_i$ is obtained by means of an $S$-matrix approach (see Appendix~\ref{AppB} for details). Moreover, the operator
$\mathbb{U}^{(21)}=(\mathds{1}-\mathbb{R}_2\mathbb{R}_1)^{-1}$
appearing in Eq.~\eqref{trans_film} describes the multiple reflection inside the cavity formed by the two films, the information about the distance being inside the scattering operators~\cite{Messina2011,Messina2014}.

The conductance $G_{bi}$ between each film and the environment can be written under the same form as Eq.~\eqref{cond_film}, with only propagative modes ($\omega>ck$) participating and with the transmission coefficient~\cite{Messina2011,Messina2014}
\begin{equation}\begin{split}
&\mathcal{T}_{bi}=\mathrm{Tr}\Bigl[\mathbb{U}^{(21)}\mathbb{T}_2\mathbb{T}_2^\dag\mathbb{U}^{(21)\dag}(\mathds{1}-\mathbb{R}_1^\dag\mathbb{R}_1-\mathbb{T}_1^\dag\mathbb{T}_1)\\
&\,+(1-\mathbb{U}^{(12)}\mathbb{T}_1\mathbb{T}_1^\dag\mathbb{U}^{(12)\dag})(\mathds{1}-\mathbb{R}_2^\dag\mathbb{R}_2)+\mathbb{R}_2\mathbb{R}_2^\dag-\mathbb{R}_{12}\mathbb{R}_{12}^\dag\Bigr],
\end{split}\label{trans_film2}\end{equation}
where $\mathbb{U}^{(12)}$ is defined as $\mathbb{U}^{(21)}$ but with $\mathbb{R}_1$ and $\mathbb{R}_2$ exchanged, while $\mathbb{R}_{12}$ is the reflection operator of the bodies 1 and 2 seen as a unique body. The transmission coefficients given in Eqs.~\eqref{trans_film} and \eqref{trans_film2} allow us to calculate the conductances $G_{12}$ and $G_b=G_{b1}=G_{b2}$, shown in Fig.~\ref{Fig_filmconductance} as a function of the applied magnetic field $H$. We first remark that $G_{12}$ is two orders of magnitude larger than $G_b$, coherently with the well-known near-field amplification of heat flux between the films. Moreover, we observe that, while the second derivative of $G_b$ is always positive, and hence $\mathscr{G} > 0$, there should be an accelerated relaxation. On the other hand, this is not the case for $G_{12}$, which has $\mathscr{G} > 0$ for field strengths smaller than $4T$ and $\mathscr{G} < 0$ for field strengths larger than 4T, allowing for an accelerated or decelerated relaxation as we will show hereafter. 

\begin{figure}
	\centering
	\includegraphics[width=0.4\textwidth]{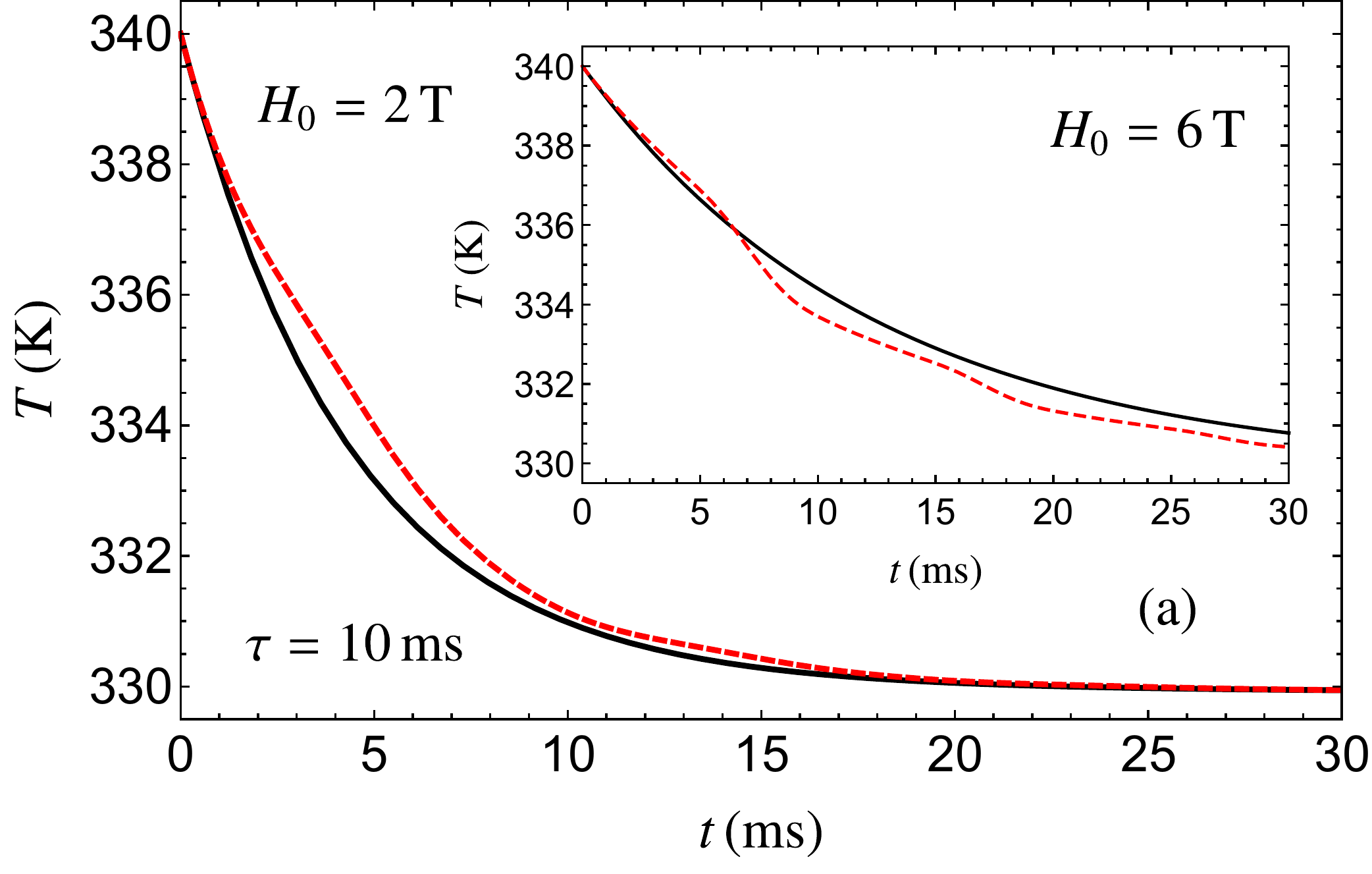}\\
	\includegraphics[width=0.4\textwidth]{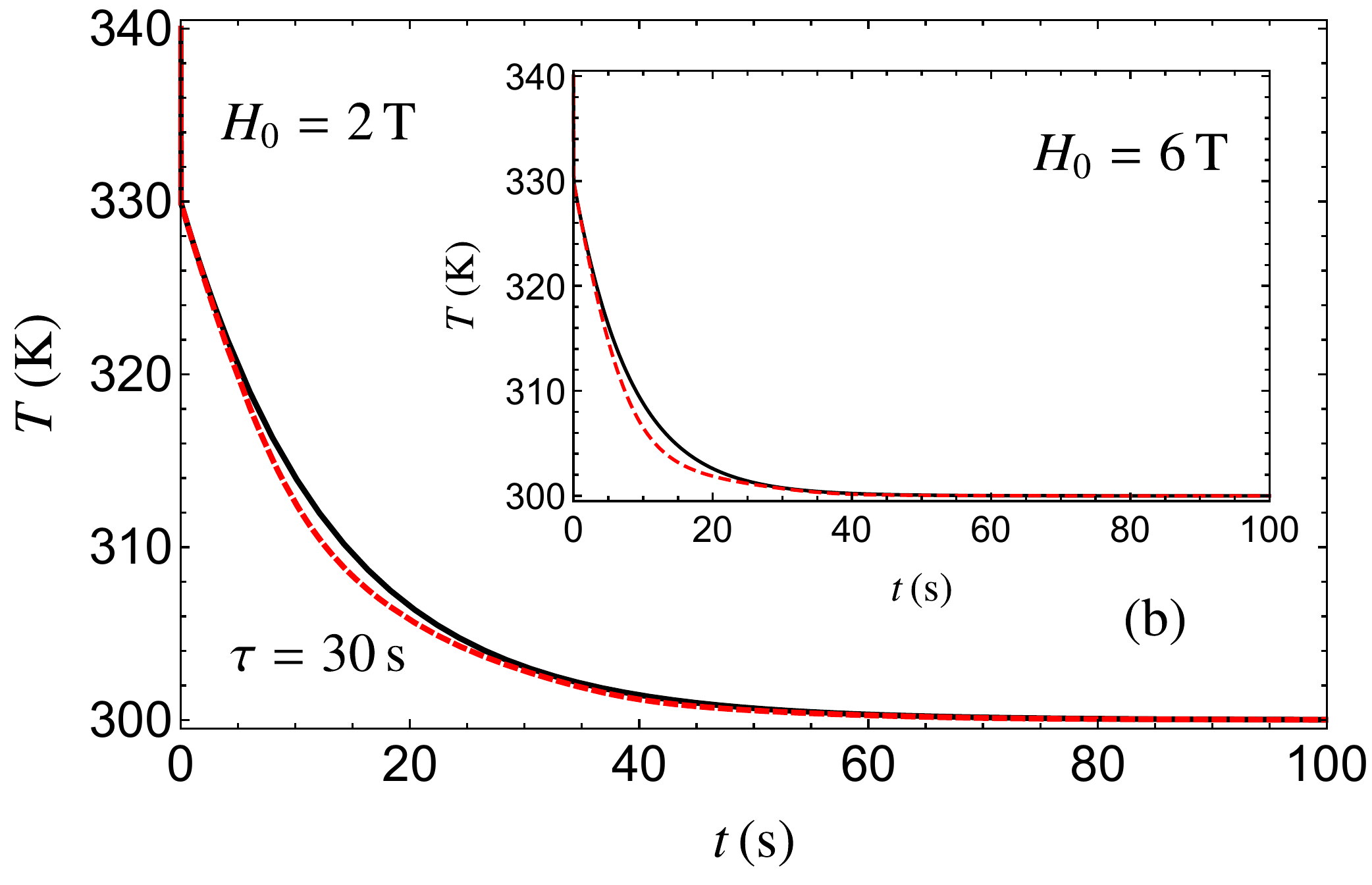}\\
	\caption{Time evolution of the temperature $T_1$ in a system made of two 1\,$\mu$m-thick InSb films placed at a distance $d=100\,$nm. In all the curves, the solid black line corresponds to a constant field $H=H_0$, while the dashed red line corresponds to the oscillating field $H(t)=H_0+\Delta H\sin(2\pi t/\tau)$. For each panel, the main part shows the time evolution for $H_0=\Delta H=2\,$T, while in the inset we have $H_0=6\,$T and $\Delta H=2\,$T. (a) Time evolution for an oscillating period $\tau=10\,$ms during the first 30\,ms (near-field thermalization between the two films). (b) Time evolution for an oscillating period $\tau=30\,$s during the first 100\,s (including thermalization with the external bath). }
	\label{Fig_filmT}
\end{figure}

By using these calculated conductances, we followed the temperature evolution starting from the intial condition $(T_1(0),T_2(0))=(340,320)\,$K. This evolution is shown in Fig.~\ref{Fig_filmT} for a field oscillation period $\tau=10\,$ms [panel (a)] and $\tau=30\,$s [panel (b)]. We show both the result for a constant field of magnitude $H_0=2\,$T and $H_0=6\,$T (solid black lines), and the one for an oscillating field $H(t)=H_0+\Delta H\sin(2\pi t/\tau)$, with $\Delta H=2\,$T (dashed red line). Consistently with the previous theoretical developements, we observe that the temporal modulation accelerates the cooling when the conductance is convex around $H_0$ and it slows down the relaxation process when it is concave [Fig.~\ref{Fig_filmT} (a)]. We observe a thermalization between the two films to a temperature $T\simeq330\,$K on a short timescale, followed by a thermalization to the bath temperature $T_b=300\,$K on a time scale around 100\,s. These different time scales, already described in Ref.~\cite{Messina2013} in the case of nanoparticles, are a direct consequence of the difference in the values of the conductances. 
\begin{figure}
	\centering
	\includegraphics[width=0.4\textwidth]{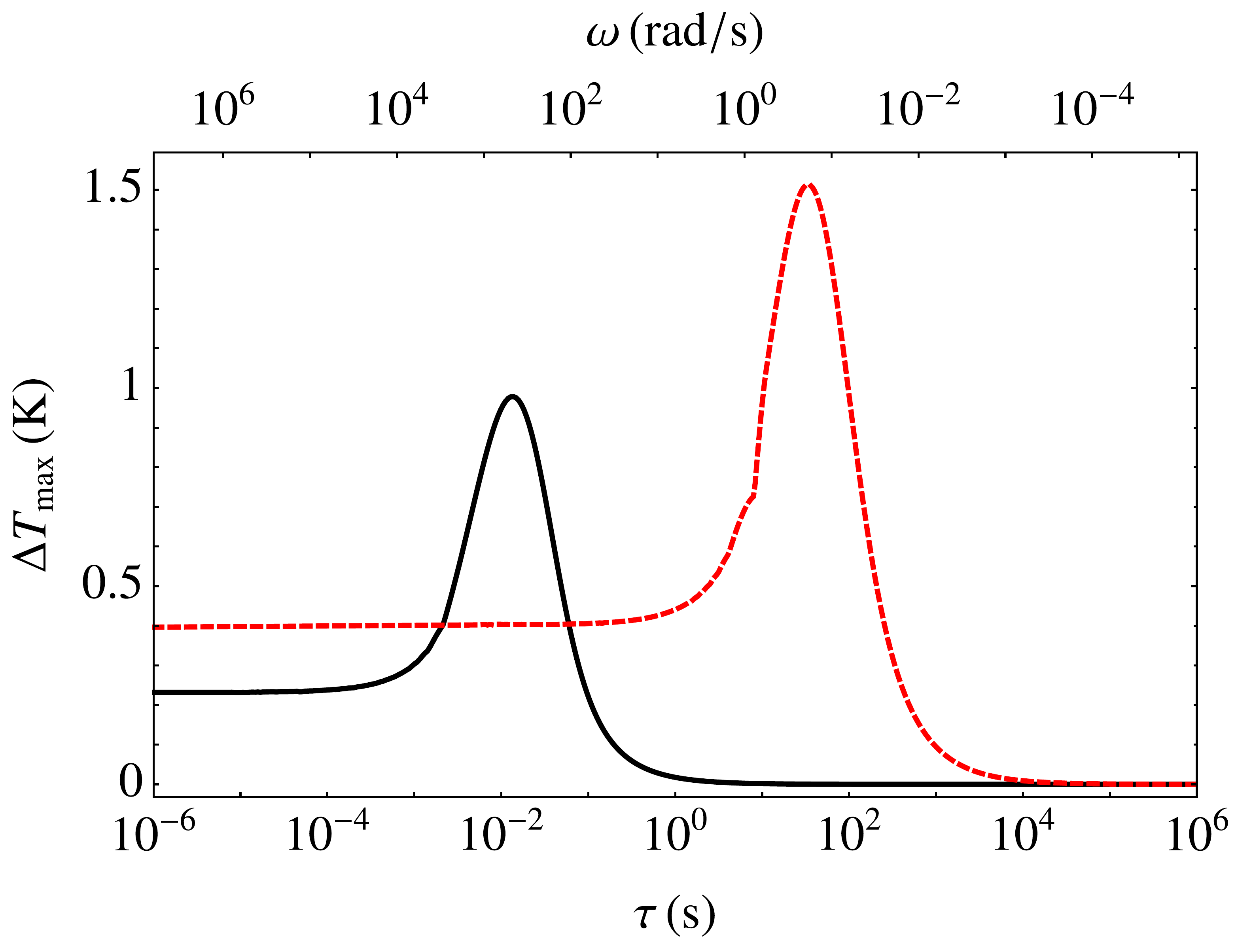}
	\caption{Maximum temperature difference (in absolute value) during the first 30\,ms of evolution (near-field thermalization between the films, solid black line) and during the rest of time evolution (far-field thermalization with the environment, dashed red line).}
	\label{Fig_filmdelta}
\end{figure}
In the time evolutions shown here, we observe a temperature difference (between constant and oscillating field), as in the case of the two particles, reaching at most a value between 1\,K and 2\,K. This confirms that the variation induced by the oscillating field exists also in a macroscopic scenario, and the effects are comparable to what we already observed for two nanoparticles.

It is interesting to address in more detail the role played by the period of the oscillating field. To this aim, we quantify the deviation induced by the field oscillation in terms of the maximum temperature difference, between the case without and with field oscillations around $H_0=2\,$T, during the first film\,-\,film thermalization ($t<30\,$ms) and during the rest of the time evolution (film-environment thermalization). We first observe that both temperature differences go to zero for very long periods $\tau$, as expected since in this case the field does not vary on the time scale of the observed evolutions. Moreover, both temperature differences are peaked at a given period $\tau$, roughly corresponding to the characteristic time scale of thermal relaxation (tens of milliseconds for the near-field interaction, tens of seconds for the bath relaxation). Finally, we note that for high frequencies of field oscillation the temperature difference converges toward a constant value (see Appendix~\ref{AppC} for more details). 

In conclusion, we have shown that the modulation of heat flux exchanged in a system out of thermal equilibrium induced by an external field can significantly alter its relaxation dynamics. This effect can be attributed to a broken time-reversal symmetry in the system. When the external field is periodically modulated a pumping effect induced by the conductance curvature over a half period is not compensated by the field modulation on the second half period. It follows a cooling effect when this modulation takes place in regions where the channels of heat exchanges are convex function of the external field. We have illustrated this mechanism in magneto-optical systems under the action of a time-modulated magnetic field. Many other fields could be used to achieve a similar effect. Hence, for instance, mechanical strains~\cite{Wang} which can be applied directly to a solid or using an external electric field thank to an inverse piezoelectric effect could be used to control the relaxation dynamics.

We have seen that the magnitude of the extracted power induced by the modulation of external field is related to the local curvature of conductances with respect to this field. Thus, a natural way to enhance the cooling effect is to increase this curvature. However in the examples we have discussed we have seen that the conductances vary smoothly with the magnetic field. But it has been shown recently that in a class of crystals a singular angular magnetoresistance can appear around specific orientations of the magnetic field~\cite{Suzuki}, leading to a very sharp variation of electric resistance with respect to field angle. The presence of singular thermal magnetoresistance could lead to a significant increase of pumping effect in periodically-driven systems.

\begin{widetext}

\appendix

\section{Fundamental matrix in linear differential systems}\label{AppA}

The fundamental matrix of system (7) in the main text at any time $t$ can be easily calculated using the resolvant $\mathcal{R}$ of system as follow
\begin{equation}
\mathbf{\phi}(t)=\underset{i=0,N-1}{\prod}\mathcal{R}(t_{i+1},t_i) \label{fund_matrix}
\end{equation}
(here the range $[0,t]$ is uniformly divided in $0=t_0<t_1<...<t_{N-1}<t=t_N$)
with the resolvant calculated using the Runge-Kutta scheme of fourth order
\begin{equation}
\mathcal{R}(t_{i+1},t_i)=\mathds{I}+\frac{\Delta t}{6} \Bigl[\mathds{C}^{-1}\mathds{G}(t_i)+2\Bigl(1-\frac{1}{\sqrt{2}}\Bigr)\mathds{E}(t_i)+2\Bigl(1+\frac{1}{\sqrt{2}}\Bigr)\mathds{F}(t_i)+\mathds{H}(t_i)\Bigr]\label{resolvant}
\end{equation}
with
\begin{equation}
\mathds{E}(t)=\mathds{C}^{-1}\mathds{G}\Bigl(t+\frac{\Delta t}{2}\Bigr)\Bigl[\mathds{I}+\frac{\Delta t}{2}\mathds{C}^{-1}\mathds{G}(t)\Bigr],
\end{equation}
\begin{equation}
\mathds{F}(t)=\mathds{C}^{-1}\mathds{G}\Bigl(t+\frac{\Delta t}{2}\Bigr)\Bigl[\mathds{I}+\Bigl(-\frac{1}{2}+\frac{1}{\sqrt{2}}\Bigr)\Delta t\,\mathds{C}^{-1}\mathds{G}(t)+\Bigl(1-\frac{1}{\sqrt{2}}\Bigr)\Delta t\,\mathds{E}(t)\Bigr]
\end{equation}
and
\begin{equation}
\mathds{H}(t)=\mathds{C}^{-1}\mathds{G}(t+\Delta t)\Bigl[\mathds{I}-\frac{1}{\sqrt{2}}\Delta t \mathds{E}(t)+\Bigl(1+\frac{1}{\sqrt{2}}\Bigr)\Delta t\,\mathds{F}(t)\Bigr].
\end{equation}

\section{Light scattering and Fresnel coefficients in non-reciprocal planar systems}\label{AppB}

We describe here the procedure to obtain the reflection and transmission operators of a film of finite thickness $\delta$ (occupying the region $[0,\delta]$ along the $z$ axis) made of a material having a dielectric permittivity tensor described by Eq.~(14) of the main text. Assuming a dependence of the fields of the form
\begin{equation}
\mathbf{E}, \mathbf{D}, \mathbf{H}, \mathbf{B} \propto e^{i\mathbf{k}\cdot\mathbf{r}}e^{iqz}e^{-i\omega t},
\end{equation}
and expressing the $z$ components $E_z$ and $H_z$ of the fields as a function of the $x$ and $y$ components, the Maxwell equations can be easily cast under the form
\begin{equation}
\mathbb{L}\begin{pmatrix}E_x\\E_y\\H_x\\H_y\end{pmatrix} = \begin{pmatrix}0\\0\\0\\0\end{pmatrix},
\label{System}\end{equation}
where
\begin{equation}
\mathbb{L}=\begin{pmatrix}
0 & - q & - \mu_0 \omega & 0\\
q & 0 & 0 & \frac{k^2}{\varepsilon_0 \varepsilon_3 \omega} - \mu_0 \omega\\
\varepsilon_0 \varepsilon_1 \omega & - i \varepsilon_0 \varepsilon_2 \omega & 0 & - q\\
i \varepsilon_0 \varepsilon_2 \omega & \varepsilon_0 \varepsilon_1 \omega - \frac{k^2}{\mu_0 \omega} & q & 0
\end{pmatrix}
\end{equation}
The system \eqref{System} has non-vanishing solutions under the condition $\det\mathbb{L}=0$, which gives $q=\pm q_{1,2}$, with
\begin{equation}\begin{split}
q_{1,2} &= \sqrt{\varepsilon_1 \frac{\omega^2}{c^2} - \frac{\varepsilon_1 + \varepsilon_3}{2 \varepsilon_3} k^2 \pm \sqrt{\frac{k^4}{4}\Bigl(\frac{\varepsilon_1 - \varepsilon_3}{\varepsilon_3}\Bigr)^2 + \frac{\varepsilon_2^2}{\varepsilon_3}\frac{\omega^2}{c^2}\Bigl(\varepsilon_3\frac{\omega^2}{c^2} - k^2\Bigr)}}.\\
\end{split}\end{equation}
For a given root $q$ of $\det\mathbb{L}=0$, the system given in Eq.~\eqref{System} can be solved under the form ($i=1,2$)
\begin{equation}
\begin{pmatrix}E_x\\E_y\\H_x\\H_y\end{pmatrix} = \begin{pmatrix}\alpha_(q_i)\\\beta(q_i)\\\gamma(q_i)\\\delta(q_i)\end{pmatrix} T_i.
\end{equation}
In the vacuum regions on the left and right side of the film, the field propagating in direction $\phi=+,-$ can be written as
\begin{equation}
\begin{pmatrix}E_x\\E_y\\H_x\\H_y\end{pmatrix} = \begin{pmatrix}-\frac{k_y}{k}\\\frac{kx}{k}\\-\frac{\phi  k_x k_z}{\mu_0 \omega k}\\-\frac{\phi k_y k_z}{\mu_0 \omega k}\end{pmatrix} A^\phi_\text{TE} + \begin{pmatrix}\frac{c\phi k_x k_z}{\sqrt{\varepsilon} \omega k}\\\frac{c\phi k_y k_z}{\sqrt{\varepsilon} \omega k}\\-\frac{\sqrt{\varepsilon} k_y}{c \mu_0 k}\\\frac{\sqrt{\varepsilon} k_x}{c \mu_0 k}\end{pmatrix} A^\phi_\text{TM},
\end{equation}
where the amplitudes of TE and TM polarizations have been explicitly identified. We now need to impose the continuity of $E_x$, $E_y$, $H_x$ and $H_y$ at the boundaries $z=0$ and $z=\delta$. We write these conditions under the form of an $S$-matrix approach (outgoing fields as a function of ingoing ones) as
\begin{equation}
\begin{pmatrix}R_\text{TE}\\R_\text{TM}\\T_1\\T_2\end{pmatrix} =\mathbb{S}_1\begin{pmatrix}I_\text{TE}\\I_\text{TM}\\T'_1\\T'_2\end{pmatrix}\quad\text{and}\quad\begin{pmatrix}T'_1\\T'_2\\T_\text{TE}\\T_\text{TM}\end{pmatrix} =\mathbb{S}_2\begin{pmatrix}T_1\\T_2\\I'_\text{TE}\\I'_\text{TM}\end{pmatrix}.
\end{equation}
In these expressions $R_p$ and $I_p$ ($p=\text{TE}, \text{TM}$) are the reflected and incident fields on the left side, $T_i$ and $T'_i$ ($i=1,2$) are the fields propagating to the right and left, respectively, inside the film, while $T_p$ and $I'_p$ are the transmitted field and the incoming field on the right side of the film. The matrices $\mathbb{S}_i$ are conveniently expressed as $\mathbb{S}_i=\mathbb{A}_i^{-1}\mathbb{B}_i$, where
\begin{equation}\begin{split}
\mathbb{A}_1 &= \begin{pmatrix}
0 & -\frac{c k_z}{\omega}  &  -\alpha(q_1) & -\alpha(q_2)\\
1 & 0 &  -\beta(q_1) & -\beta(q_2)\\
\frac{k_z}{\mu_0\omega}  & 0 &  -\gamma(q_1) & -\gamma(q_2)\\
0 & \frac{1}{\mu_0c}  &  -\delta(q_1) & -\delta(q_2)\end{pmatrix},\quad
\mathbb{B}_1 = \begin{pmatrix}
0 & -\frac{c k_z}{\omega}  &  \alpha(q_1) & \alpha(q_2)\\
-1 & 0 &  \beta(q_1) & \beta(q_2)\\
\frac{k_z}{\mu_0\omega}  & 0 &  -\gamma(q_1) & -\gamma(q_2)\\
0 & -\frac{1}{\mu_0c}  &  -\delta(q_1) & -\delta(q_2)\end{pmatrix},\\
\mathbb{A}_2 &= \begin{pmatrix}
\alpha(q_1) & \alpha(q_2) & 0 & -\frac{c k_z}{\omega} \\
\beta(q_1) & \beta(q_2) & -1 & 0\\
-\gamma(q_1) & -\gamma(q_2) & \frac{k_z}{\mu_0\omega}  & 0 \\
-\delta(q_1) & -\delta(q_2) & 0 & -\frac{1}{\mu_0c}\end{pmatrix},\quad
\mathbb{B}_2 = \begin{pmatrix}
-\alpha(q_1) & -\alpha(q_2) & 0 & -\frac{c k_z}{\omega} \\
-\beta(q_1) & -\beta(q_2) & 1 & 0\\
-\gamma(q_1) & -\gamma(q_2) & \frac{k_z}{\mu_0\omega}  & 0 \\
-\delta(q_1) & -\delta(q_2) & 0 & \frac{1}{\mu_0c}\end{pmatrix}.
\end{split}\end{equation}
We finally deduce the full $\mathbb{S}$ matrix of the system, defined by the relation
\begin{equation}
\begin{pmatrix}R_\text{TE}\\R_\text{TM}\\T_\text{TE}\\T_\text{TM}\end{pmatrix} =\mathbb{S}\begin{pmatrix}I_\text{TE}\\I_\text{TM}\\I'_\text{TE}\\I'_\text{TM}\end{pmatrix}=\begin{pmatrix}\mathbb{S}_{11} & \mathbb{S}_{12}\\\mathbb{S}_{21} & \mathbb{S}_{22}\end{pmatrix}\begin{pmatrix}I_\text{TE}\\I_\text{TM}\\I'_\text{TE}\\I'_\text{TM}\end{pmatrix},
\end{equation}
and given by $\mathbb{S}=\mathbb{S}_1\circledast\mathbb{S}_2$, having introduced the associative operation $\mathbb{A}=\mathbb{B}\circledast\mathbb{C}$,  defined as
\begin{equation}\begin{split}
\mathbb{A}_{11}&=\mathbb{B}_{11}+\mathbb{B}_{12}(\mathds{1}-\mathbb{C}_{11}\mathbb{B}_{22})^{-1}\mathbb{C}_{11}\mathbb{B}_{21},\\
\mathbb{A}_{12}&=\mathbb{B}_{12}(\mathds{1}-\mathbb{C}_{11}\mathbb{B}_{22})^{-1}\mathbb{C}_{12},\\
\mathbb{A}_{21}&=\mathbb{C}_{21}(\mathds{1}-\mathbb{B}_{22}\mathbb{C}_{11})^{-1}\mathbb{B}_{21},\\
\mathbb{A}_{22}&=\mathbb{C}_{22}+\mathbb{C}_{21}(\mathds{1}-\mathbb{B}_{22}\mathbb{C}_{11})^{-1}\mathbb{B}_{22}\mathbb{C}_{12}.
\end{split}\end{equation}
We conclude by identifying the $\mathbb{S}_{11}$ block with the reflection matrix of the film, and the $\mathbb{S}_{21}$ block with the transmission matrix. As discussed for example in Refs.~[20,21], the final step to be done is the introductiong of phase factors multiplying the scattering operators in order to take into account the coordinate of each body.

\section{Maximum temperature difference due to the presence of an oscillating field}\label{AppC}

We focus here on the discussion of the behavior of the curves shown in Fig.~4 of the main paper for high frequency (low period). We have seen that in this limit the maximum temperature difference between the configurations with and without field oscillation goes to a constant value in this limit, different from zero. In this section, we briefly discuss this behavior in the case of a single body thermalizing with an environment. In the scenario, studied in the paper, of a field oscillating only in one direction, the time evolution of the temperature $T(t)$ is described by the equation
\begin{equation}
C\frac{dT}{dt}=G[F(t)] (T_b-T),
\end{equation}
having solution
\begin{equation}
T(t)=T_b+(T(0)-T_b)\exp\Bigl[-\frac{1}{C}\int_{0}^{t}G[F(\tau)] d\tau\Bigr].\label{Solution}
\end{equation}
If we consider an external field $F(t)=F_0+\Delta F\sin(2\pi t/T)$, oscillating with period $T$ around a constant value $F_0$, the conductance can be expanded around $F_0$ as follows
\begin{equation}
G[F(t)] = G(F_0)+\sum_{n=1}^{\infty}\frac{G^{(n)}(F_0)}{n!}(\Delta F)^n\sin^n\Bigl(\frac{2\pi t}{T}\Bigr),
\end{equation}
where $G^{(n)}(F_0)$ is the $n$-th derivative of $G$ calculated at $F_0$. The integral appearing in Eq.~\eqref{Solution} can be thus written as
\begin{equation}
\int_{0}^{t}d\tau\,G[F(\tau)] = G(F_0)t+\sum_{n=1}^{\infty}\frac{G^{(n)}(F_0)(\Delta F)^n}{n!}\int_{0}^{t}d\tau\sin^n\Bigl(\frac{2\pi\tau}{T}\Bigr).
\label{Integral}\end{equation}
At this stage, we assume that $T$ is sufficiently small, compared to any $t$ at which we are interested in knowing $T(t)$, to allow us to perform the approximation $t=nT$ for some integer $n$ and thus
\begin{equation}
\int_{0}^{t}d\tau\sin^n\Bigl(\frac{2\pi\tau}{T}\Bigr)=\begin{cases}
\frac{2t\Gamma\bigl(\frac{n+1}{2}\bigr)}{n\sqrt{\pi}\Gamma\bigl(\frac{n}{2}\bigr)} & n\,\text{even}\\
0 & n\,\text{odd}
\end{cases}
\end{equation}
This allows us to rewrite the integral \eqref{Integral} as
\begin{equation}
\int_{0}^{t}d\tau\,G[F(\tau)] = \bigl[G(F_0)+\mathscr{G}(F_0,\Delta F)\bigr]t,
\end{equation}
being
\begin{equation}
\mathscr{G}(F_0,\Delta F) = \sum_{n=1}^{\infty}\frac{G^{(2n)}(F_0)(\Delta F)^{2n}\Gamma\bigl(n+\frac{1}{2}\bigr)}{\sqrt{\pi}\,n!(2n)!}= \sum_{n=1}^{\infty}\frac{G^{(2n)}(F_0)(\Delta F)^{2n}}{4^nn^2[(n-1)!]^2}.
\label{Development}\end{equation}
We remark that the solution obtained is clearly independent of $T$ and corresponds to the high-frequency limit. We have thus proved that in the high-frequency limit the time behavior of the temperature is still exponential. The temperature difference with and without field oscillation reads
\begin{equation}
\Delta T(t)=(T(0)-T_b)\exp\Bigl[-\frac{G(F_0)}{C}t\Bigr]\Bigl\{1-\exp\Bigl[-\frac{\mathscr{G}(F_0,\Delta F)}{C}t\Bigr]\Bigr\}.
\end{equation}
As expected, this temperature difference vanishes both at $t=0$ and for $t\to\infty$. This quantity is maximized at
\begin{equation}
t_\text{max} = \frac{C}{\mathscr{G}(F_0,\Delta F)}\log\Bigl(1 + \frac{\mathscr{G}(F_0,\Delta F)}{G(F_0)}\Bigr),
\end{equation}
where it takes the value
\begin{equation}
\Delta T_\text{max}=(T(0)-T_b)\frac{\mathscr{G}(F_0,\Delta F)}{G(F_0)}\Bigl(\frac{G(F_0)}{G(F_0) + \mathscr{G}(F_0,\Delta F)}\Bigr)^{\frac{G(F_0) + \mathscr{G}(F_0,\Delta F)}{\mathscr{G}(F_0,\Delta F)}}.
\end{equation}
We conclude by highlighting that this temperature difference is independent of $C$ and proportional to the initial temperature difference $T(0)-T_b$. This expression can be used to obtain an estimate of the impact of the field oscillation without solving the entire dynamics of the problem.

This expression can be exploited to predict the maximum temperature difference in the high frequency limit obtained numerically in Fig.~\ref{Fig_filmdelta}. To this aim, we start with a parabolic fit for the conductance $G_b$ (see Fig.~\ref{Fig_filmconductance}). This allows to keep only one term ($n=1$) in Eq.~\eqref{Development}. Finally, we choose $T(0)=330\,$K, corresponding to the temperature of near-field thermalization between the two films (see Fig.~\ref{Fig_filmT}) and obtain $\Delta T_\text{max}=0.28\,$K, in good agreement (within the assumption of parabolic behavior) with the high-frequency limit highlighted in Fig.~\ref{Fig_filmdelta}.

\end{widetext}

\end{document}